# BRAIN ACTIVITY ON A HYPERSPHERE

**Arturo Tozzi**

*Center for Nonlinear Science, University of North Texas*
*1155 Union Circle, #311427*
*Denton, TX 76203-5017 USA*
tozziarturo@libero.it

**James F. Peters**

*Department of Electrical and Computer Engineering, University of Manitoba*
*75A Chancellor's Circle*
*Winnipeg, MB R3T 5V6 CANADA*
James.Peters3@umanitoba.ca

*Current advances in neurosciences deal with the functional architecture of the central nervous system, paving the way for "holistic" theories that improve our understanding of brain activity. From the far-flung branch of topology, a strong concept comes into play in the understanding of brain signals, namely, continuous mapping of the signals onto a "hypersphere": a 4D space equipped with a donut-like shape undetectable by observers living in a 3D world. Here we show that the brain connectome may be regarded as a functional hypersphere. We evaluated the features of the imperceptible fourth dimension based on resting-state fMRI series. In particular, we looked for simultaneous activation of antipodal signals on the 3D cortical surface, which is the topological hallmark of the presence of a hypersphere. Here we demonstrate that spontaneous brain activity displays the typical features which reveal the existence of a functional hypersphere. We anticipate that this introduction to the brain hypersphere is a starting point for further evaluation of a nervous' fourth spatial dimension, where mental operations take place both in physiological and pathological conditions. The suggestion here is that the brain is embedded in a hypersphere, which helps solve long-standing mysteries concerning our psychological activities such as mind-wandering and memory retrieval or the ability to connect past, present and future events.*

An *n*-sphere is a n-dimensional structure that is a generalization of a circle. Specifically, an n-sphere with radius R is set of n-tuples of points. For example, a 2-sphere is a set points on the perimeter of a circle in a 2D space, a 3-sphere is a sets of surface points in a 3D space (a beach ball is a good example) and a 4-sphere is set of points on the surface of what is known as a hypersphere. The prefix "hyper" refers to 4- (and higher-) dimensional analogues of 3D spheres. In mathematical terms, a 4-sphere, also called *glome* or generically *hypersphere*, is a simply connected manifold of constant, positive curvature, enclosed in an Euclidean 4-dimensional space called a *4-ball*. The term glome comes from the Latin "glomus", meaning ball of string. A 4-sphere is thus the surface or boundary of a 4-dimensional ball, while a 4-dimensional ball is the interior of a 4-sphere. A glome can be built by superimposing two 3-spheres whose opposite edges are abstractly glued together: we obtain a topological structure, the *Clifford torus*. A Clifford torus is a special kind of torus (donut shape) that is a minimal surface which sits inside a glome and is equipped with intricate rotations, called *quaternionic movements* (**Figure 1**). Such a torus has the same local geometry as an "ordinary" three-dimensional space, but its global topology is different. The hypersphere, requiring four dimensions for its definition just as an ordinary sphere requires three, is not detectable in the usual spatial 3-dimensions and is thus challenging to assess. **Figure 1** shows the possible ways to cope with a 3D visualization of a glome. In this paper, we hypothesize that brain activity is shaped in guise of an hypersphere which performs 4D movements on the cortical layers, giving rise to a functional Clifford torus where mental operations take place.

Experimental and theoretical clues allow us to conjecture that the brain activities (at least some of them) are embedded in a torus lying on the surface of a hypersphere. The theoretical claims of brain multidimensionality are widespread *(1-3)* and models characharcterized by dimensionality reduction have been used in the study of human central nervous system *(4)*. It has been demonstrated that spontaneous activity structures of high dimensionality – termed "lag threads" - can be found in the brain, consisting of multiple highly reproducible temporal sequences *(5)*. Recent findings suggest that nervous structures process information through topological as well as spatial mechanisms. For example, it is has been hypothesized that hippocampal place cells create topological templates to represent spatial information *(6)*. The glome displays a double-torus shape, *i.e.*, the trajectory followed by a particle inside the torus is closed and similar to a video game with biplanes in aerial combat. When a biplane flies off one edge of gaming display, it does not crash but rather

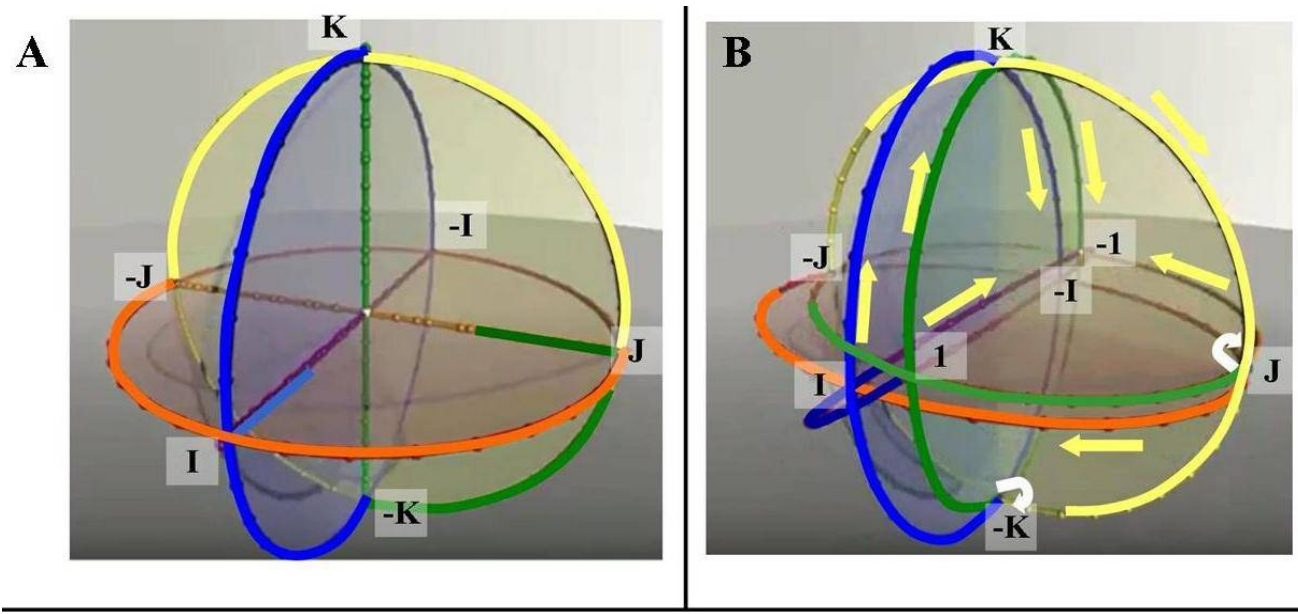

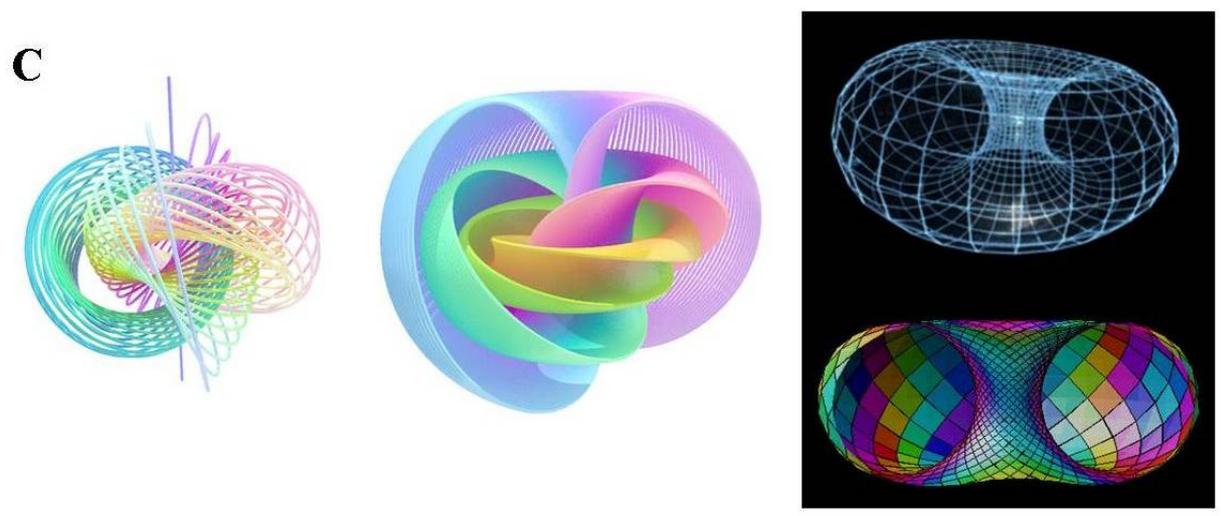

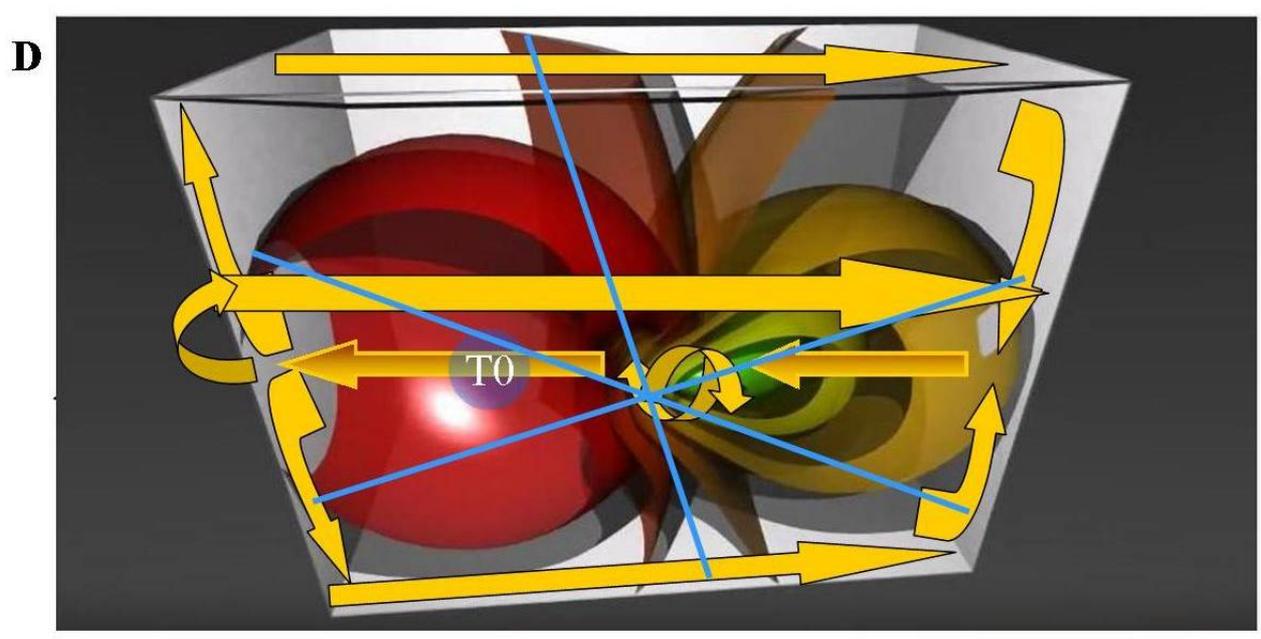

**Figure 1.** Different ways to depict a hypersphere. To better understand the concept of a 4-sphere, the images should be watched during their complicated movements, *i.e.*, it is helpful to watch the videos mentioned below.

**Figure 1A** shows the three circumferences embedding a "normal" 3-sphere in a 3D space equipped with the "classical" 3D coordinates.

**Figure 1B** shows how the superimposition of another 3-sphere (which circumference is glued together with the one of the sphere of **Figure 1A**) gives rise to a 3-sphere (from video https://www.youtube.com/watch?v=XFW769hqa1U ). Each apparent line segment is really two line segments, one arching upward into the third dimension and the other arching downward. Observe how opposite sections of the rim fit together, rather than trying to visualize the whole thing at once the way you would visualize a sphere. The four pairs of antipodal points (called *1,-1, k, -k, j,-j* and *i, -i*) give rise to the so-called "quaternion group", equipped with two possible types of reciprocal 4D rotations. Some of the quaternion rotations are depicted, as an example, by the yellow and white arrows.

**Figure 1C** shows how a glome can be formed by different circles arranged in 4D (right panel). The shape of the glome is everchanging, depending on the number of circles taken into account and their trajectories (see the video at http://nilesjohnson.net/hopf.html). The central and left panels show another way to depict a hypersphere: two spheres glued together along their spherical boundary give rise to a Clifford torus (https://en.wikipedia.org/wiki/Clifford_torus#/media/File:Clifford-torus.gif shows a stereographic projection of a Clifford torus performing a simple rotation through the *xz* plane).

**Figure 1D** shows the 3D Stereographic projection of the "toroidal parallels" of a glome (from https://www.youtube.com/watch?v=QlcSlTmc0Ts ; see also http://www.matematita.it/materiale/index.php?lang=en&p=cat&sc=2,745). The orange arrows illustrate the trajectories followed by the 4D quaternionic movements of a Clifford torus when projected onto the surface of the 3D space in which it is embedded. Note that the arrows follow the external and medial surfaces of the 3D space in a way that is predictable. Just one of the possible directions of the quaternion movements is displayed: the flow on a Clifford torus may occur in each of the four planes. In this case, the spheres on the left increase in diameter, forming a circle of increasing circumference on the left surface of the 3D space. Conversely, on the opposite right side, the spheres shrink and give rise to a circle of decreasing circumference on the right surface of the 3D space. The blue lines depict some of the possible antipodal points predicted by the Borsuk Ulam Theorem. To give another example, *J* and *-J* are antipodal points in **Figure 1B**.

---

it comes back from the opposite edge of the screen *(7)*. Mathematically speaking, the display edges have been "glued" together. The human brain exhibits similar behavior, *i.e.*, the unique ability to connect past, present and future events in a single, coherent picture *(8,9)*, as if we were allowed to watch the three screens of past-present-future glued together in a mental kaleidoscope. The same occurs during other brain functions, *e.g.*, memory retrieval, recursivity of imagination and mind wandering *(10)*, in which concepts appear to be "glued" together, flowing from a state to another. The torus is naturally visualized intrinsically, by ignoring any extrinsic properties a surface may have (it is thought that all the movements onto a torus surface are performed just by trajectories internal to its structure). For example, take a sheet of paper and bend it into a half-cylinder. The extrinsic geometry of the paper has obviously changed, but the paper itself has not been deformed and its intrinsic geometry has not varied. What would you see if you lived in a closed three-manifold? You should be able to see yourself, via the intrinsic structure provided by the glued surfaces of a hypersphere, in an otherwise unperceivable 4D space[7]. In the same way, we humans perceive our thoughts intrinsically and naturally adopt "private", subjective standpoints.

## MATERIALS AND METHODS

**The movements of particles on a glome.**

At first, we need to mathematically define a hypersphere *(11,12)*. It is an n-sphere formed by points which are constant distance from the origin in $(n+1)$-dimensions. A 4-sphere (also called glome) of radius *r* (where r may be any positive real number) is defined as the set of points in 4D Euclidean space at distance *r* from some fixed center point **c** (which may be any point in the 4D space).

In technical terms, in our study we projected onto a 3-D surface a map of a glome equipped with Sp(1) or SU(2) Lie groups. The 4-sphere is parallelizable as a differentiable manifold, with a principal U(1) bundle over the 3-sphere. The only other spheres that admit the structure of a Lie group are the 0-sphere $S^0$ (real numbers with absolute value 1), the circle $S^1$ (complex numbers with absolute value 1), $S^3$, and $S^7$.

The 4-sphere's Lie group structure is Sp(1), which is a compact, simply connected symplectic group, equipped with with $\frac{\dim}{R} = 1(2 \times 1 + 1) = 3$ and quaternionic $1 \times 1$ unitary matrices. Indeed, the glome $S^4$ forms a Lie group by identification with the set of quaternions of unit norm, called versors *(13)*. The quaternionic manifold is a cube with each face glued to the opposite face with a one quarter clockwise turn. The name arises from the fact that its symmetries can be modelled in the quaternions, a number system like the complex numbers but with three imaginary quantities, instead of just one *(14)*. For an affordable, less technical treatment of quaternions, see *(15)* and the correlated, very useful video: http://blogs.scientificamerican.com/roots-of-unity/nothing-is-more-fun-than-a-hypercube-of-monkeys/ .
In addition: Sp(1) ≈ SO(4)/SO(3)≈Spin(3)≈SU(2).
Thus, Sp(1) is equivalent to - and can be identified with - the special unitary group SU(2).

**The Borsuk-Ulam Theorem.**

Brains equipped with a hypersphere is a counter-intuitive hypothesis, since we live in a 3D world with no immediate perception that 4D space exists at all, *e.g.*, if you walk along one of the curves of a 4-ball, you think are crossing a straight trajectory, and do not recognize that your environment is embedded in higher dimensions. We need to evaluate indirect clues of the undetectable fourth dimension, such as signs of the glome rotations on a familiar 3D surface. In other words, rotations of a torus embedded in a 4-ball can be identified through their "cross section" movements on a more accessible 3D surface (**Figure 1D**), as if you recognized an object from its shadow projected on a screen. The presence of a glome can be detected invoking the Borsuk-Ulam Theorem (BUT), which states that every continuous map from a hypersphere to a 3D Euclidean space must identify a pair of antipodal points (i.e., points directly opposite each other) (**Figure 1D**). This leads naturally to the possibility of a region-based, instead of a point-based, geometry in which we view collections of signals as surface shapes, where one shape maps to another antipodal one.

Continuous mappings from object spaces to feature spaces lead to various *incarnations* of the Borsuk-Ulam Theorem, a remarkable finding about Euclidean n-spheres and antipodal points by K. Borsuk *(16)*. Briefly, antipodal points are points opposite each other on a circle or on what is known as an n-sphere (called *hypersphere*). There are natural ties between Borsuk's result for antipodes and mappings called homotopies. The early work on n-spheres and antipodal points eventually led Borsuk to the study of retraction mappings and homotopic mappings *(17-19)*.
The Borsuk-Ulam Theorem states that:

*Every continuous map $f : S^n \to R^n$ must identify a pair of antipodal points*.

The notation $S^n$ denotes an n-sphere, which is a generalization of the circle. From a geometer's perspective, we have the following n-spheres, starting with the perimeter of a circle (this is $S^2$) and advancing to $S^4$, which is the smallest hypersphere.

2-sphere $S^2 : x_1^2 + x_2^2 \to R^2$ (circle perimeter),
3-sphere $S^3 : x_1^2 + x_2^2 + x_3^2 \to R^2$ (surface),
4-sphere $S^4 : x_1^2 + x_2^2 + x_3^2 + x_4^2 \to R^2$ (smallest hypersphere surface), ...,
n-sphere $S^n : x_1^2 + x_2^2 + x_3^2 + .... + x_n^2 \to R^2$.

Points are *antipodal*, provided the points are diametrically opposite *(20)*. Examples are the endpoints of a line segment or opposite points along the circumference of a circle, or poles of a sphere. An *n*-dimensional Euclidean vector space is denoted by $R^n$. In terms of brain activity, a feature vector $x \in R^n$ models the description of a brain signal.

To complete the picture in the application of the Borsuk-Ulam Theorem in brain signal analysis, we view the surface of the brain as a sphere and the feature space for brain signals as finite Euclidean topological spaces. The Borsuk-Ulam Theorem tells us that for description $f(x)$ for a brain signal $x$, we can expect to find an antipodal feature vector $f(-x)$ that describes a brain signal on the opposite (antipodal) side of the brain. Moreover, the pair of antipodal brain signals have matching descriptions.

Let $X$ denote a nonempty set of points on the surface of the brain. A topological structure on $X$ (called a brain topological space) is a structure given by a set of subsets $\tau$ of $X$, having the following properties:

(Str.1) Every union of sets in $\tau$ is a set in $\tau$
(Str.2) Every finite intersection of sets in $\tau$ is a set in $\tau$

The pair $(X, \tau)$ is called a topological space. Usually, $X$ by itself is called a topological space, provided $X$ has a topology $\tau$ on it. Let $X, Y$ be topological spaces. Recall that a function or map $f : X \to Y$ on a set $X$ to a set $Y$ is a subset $X \times Y$ so that for each $x \in X$ there is a unique $y \in Y$ such that $(x, y) \in f$ (usually written $y = f(x)$). The mapping $f$ is defined by a rule that tells us how to find $f(x)$. For a good introduction to mappings, see *(21)*.

A mapping $f : X \to Y$ is continuous, provided, when $A \subset Y$ is open, then the inverse $f^{-1}(A) \subset X$ is also open. For more about this, see *(22)*. In this view of continuous mappings from the brain signal topological space $X$ on the surface of the brain to the brain signal feature space $R^n$, we can consider not just one brain signal feature vector $x \in R^n$, but also mappings from $X$ to a set of brain signal feature vectors $f(X)$. This expanded view of brain signals has interest, since every connected set of feature vectors $f(X)$ has a shape. The significance of this is that brain signal shapes can be compared.

A consideration of $f(X)$ (set of brain signal descriptions for a region $X$) instead of $f(x)$ (description of a single brain signal $x$) leads to a region-based view of brain signals. This region-based view of the brain arises naturally in terms of a comparison of shapes produced by different mappings from $X$ (brain object space) to the brain feature space $R^n$. An interest in continuous mappings from object spaces to feature spaces leads into homotopy theory and the study of shapes.

Let $f, g : X \to Y$ be continuous mappings from $X$ to $Y$. The continuous map $H : X \times [0,1] \to Y$ is defined by
$H(x,0) = f(x), \quad H(x,1) = g(x),$ for every $x \in X$.

The mapping $H$ is a *homotopy*, provided there is a continuous transformation (called a deformation) from $f$ to $g$. The continuous maps $f, g$ are called homotopic maps, provided $f(X)$ continuously deforms into $g(X)$ (denoted by $f(X) \to g(X)$). The sets of points $f(X)$, $g(X)$ are called shapes. For more about this, see *(23,24)*.

For the mapping $H : X \times [0,1] \to R^n$, where $H(X,0)$ and $H(X,1)$ are *homotopic*, provided $f(X)$ and $g(X)$ have the same shape. That is, $f(X)$ and $g(X)$ are homotopic, provided
$\|f(X) - g(X)\| < \|f(X)\|$, for all $x \in X$.

It was Borsuk who first associated the geometric notion of shape and homotopies. This leads into the geometry of shapes and shapes of space *(25)*. A pair of connected planar subsets in Euclidean space $R^2$ have equivalent shapes, provided the planer sets have the same number of holes *(22)*. For example, the letters e, O, P and numerals 6, 9 belong to the same equivalence class of single-hole shapes. In terms of brain signals, this means that the connected graph for $f(X)$ with, for example, an *e* shape, can be deformed into the 9 shape.

This suggests yet another useful application of Borsuk's view of the transformation of shapes, one into the other, in terms of brain signal analysis. Sets of brain signals not only will have similar descriptions, but also dynamic character. Moveover, the deformation of one brain signal shape into another occurs when they are descriptively near *(26)*.

**Brain activity and hyperspheres**.

In the last paragraphs we have developed a mathematical model of antipodal points and regions casted in a biologically informed fashion, resulting in a framework that has the potential to be operationalized and assessed empirically. To evaluate a hypersphere in terms of a framework for brain activity, we first need to identify potential brain signal loci where quaternion rotations might take place. The natural candidate is the spatially embedded network of the human connectome *(27)*, a non-stationary, highly dynamical structure *(28,29)* characterized by complex topological features and an ever-changing geometry *(30)* (**Fig. 2A**). We embedded the brain in the 3D space of a Clifford torus and looked on cortical surfaces for antipodal points or shapes (**Fig. 2B**). The antipodal points evoked by BUT were viewed as brain signals opposite each other on a glome, *i.e.*, when a brain surface is activated, we identified the simultaneous activation of antipodal surface signals as a proof of a perceivable "passing through" of the fourth dimension onto the brain 3D surface. The main benefit here is that, according to the BUT dictates, for each given brain signal we can find a counterpart in the antipodal positions on the cortical surface.

We have corroborated our brain hypersphere hypothesis with published resting-state fMRI data. We evaluated movies or Figures from 14 available experimental studies and/or metaanalyses describing the brain spontaneous activity, looking for the hallmarks of the hypothesized BUT.

**Which studies did we evaluate, and why?**

Spontaneous oscillations are intrinsic, low-frequency fluctuations of cerebral activity which cannot be attributed to the experimental design or other explicit input or output *(31)*. Among the networks exhibiting coherent fluctuations in spontaneous activity, the "default-mode network" (DMN) is worth of mentioning, because it includes functionally and structurally connected regions that show high metabolic activity at rest, but deactivate when specific goal-directed behavior is needed *(32)*. Spontaneous oscillations recapitulate the topographies of fMRI responses to a wide variety of sensory, motor and cognitive task paradigms, providing a powerful means of delineating brain functional organization without the need for subjects to perform tasks *(33)*.

We favoured studies focused on intrinsic, instead of task-evoked activity, because the former is associated with mental operations that could be attributed to the activity of a glome - "screens" are glued together and the trajectories of particles (or thoughts!) follow the internal surface of a Clifford torus -. For example, spontaneous brain activity has been associated with mind-wandering or day dreaming propensities *(34)*, construction of coherent mental scenes, autobiographical memories *(35)*, experiences focused on the future (for a description of the terminology, see *10*) and dreaming state *(36)*. Recent evidence also suggests overlap between the DMN and regions involved in self- and other-related mental operations – such as affective and introspective processes *(37-39)*. It has been hypothesized that spontaneous functional connectivity patterns at rest might constitute a "signature of consciousness", reflecting a stream of ongoing cognitive processes *(40)*. It has also been proposed that spontaneous activity is highly variable among individuals, depending on local brain differences, somatosensory awareness, age span, race, culture and so on *(41,42)*. We speculate that such variably might be correlated with those differences in Clifford torus' structure and movements illustrated in the main text. A brain glome has the potential to constitute a conceptual bridge, because it exhibits both anatomical/functional (spontaneous brain activity and DMN) and psychological correlates (spontaneous, deliberate, self-generated thoughts).

The images and movies we examined were extrapolated from the following papers:

a) Ajilore, O. et al. *Constructing the resting state structural connectome. Front. Neuroinform.* 7:30 (2013).
b) Andrews-Hanna, J.R. et al. *The default network and self-generated thought: component processes, dynamic control, and clinical relevance. Ann. N. Y. Acad. Sci.* **1316**, 29-52 (2014).
c) Barttfeld, P. et al. *Signature of consciousness in the dynamics of resting-state brain activity. Proc. Natl. Acad. Sci. U. S. A.* **112**, 887-892 (2015).
d) Fox, M.D., Raichle, M.E. *Spontaneous fluctuations in brain activity observed with functional magnetic resonance imaging. Nat. Rev. Neurosci.* **8**, 700-711 (2007).
e) Fox, K.C. et al. *The wandering brain: meta-analysis of functional neuroimaging studies of mind-wandering and related spontaneous thought processes. Neuroimage* **111**, 611-621 (2015).
f) Gravel, N. et al. *Cortical connective field estimates from resting state fMRI activity. Front. Neurosci.* 8: 339 (2014).
g) Gusnard, D.A. et al. *Medial prefrontal cortex and self-referential mental activity: relation to a default mode of brain function. Proc. Natl. Acad. Sci. U. S. A.* **98**, 4259-4264 (2001).
h) Harrison, S.J. et al.. *Large-scale probabilistic functional modes from resting state fMRI. Neuroimage* **109**, 217-231 (2015).
i) Karahanoglu, F.I., Van De Ville, D. *Transient brain activity disentangles fMRI resting-state dynamics in terms of spatially and temporally overlapping networks. Nat. Commun.* 6:7751 (2015).
j) Liu, X. et al. *Decomposition of spontaneous brain activity into distinct fMRI co-activation patterns. Front. Syst. Neurosci.* 7:101 (2013).
k) Mao, D, et al. *Low-Frequency Fluctuations of the Resting Brain: High Magnitude Does Not Equal High Reliability. PLoS One* 10(6):e0128117 (2015).
l) Mitra, A. et al. *Lag threads organize the brain's intrinsic activity. Proc. Natl. Acad. Sci. U. S. A.* **112**, E2235-2244. (2015).
m) Power, J.D. et al. *Studying brain organization via spontaneous fMRI signal. Neuron.* **84**, 681-696 (2014).
n) Raichle, M.E. *A paradigm shift in functional brain imaging. J. Neurosci.* **29**, 12729-12734 (2009).

**RESULTS**

We found that all the analyzed temporal series displayed the predicted signs. The whole fMRI sequences of brain region activations, apart from differences depending on slight methodological distinctions among studies, exhibited a stereotyped topographical pattern of activity, such that brain loci are activated together with their antipodal points (**Figure 2C and 3**). We found highly reproducible topography and propagation through subsets of regions that are shared across multiple trajectories: it corroborates the predictions of BUT and brain hypersphere. Brain activity is temporally driven by a functional glome, intrinsic to the brain and (probably) embedded in the very anatomical structure of the connectome. A 4D cap surrounds the brain, equipped with trajectories following quaternion rotations along the nodes of the connectome.

**CONCLUSIONS**

Our study uncovered ample evidence of hypersphere in experimental fMRI series obtained during spontaneous activity, raising the possibility that brain activity lies on a glome, embedded in 4D space. The reproducibility of the BUT hallmarks suggests that this organizational feature is essential to normal brain physiology and function. Further studies are needed to evaluate what happens when other other techniques are used, *e.g.*, EEG and diffusion tensor imaging. Does evoked, task-related activity exhibit the same features? Further investigations will elucidate whether, following the stimulus onset, the multidimensional space outlined by cortical activity is invariant or reduced *(2,3)*. Because neighboring images of the same object are related by glide reflections translations *(7)*, it remains to be seen what the implications of the hypersphere would have for consciousness, perception of time and the nature of reality. Our "deterministic" account of linear transformations needs to be contextualized, taking into account the suggestions of the brain as an energetic, complex, nonlinear system equipped with attractors and random walks *(43-45)*. The role of electromagnetic currents needs to be re-evaluated, *i.e.*, do such currents contain the message, or, as recently suggested *(46)* do they serve other kinds of functions? For example, it has been proposed that features of a brain signal with spectral peaks in preferred bands (gamma, beta and so on) provide a basis for feature vectors in a 4D euclidean space *(47)*. Further, the hypersphere, due to different transformations of the quaternionic group, continuously changes its intrinsic structure. In this context, it is reasonable to speculate that each mental state corresponds to a different glome topological space.

## ACKNOWLEDGEMENTS

The Authors would like to thank Norbert Jausovec for his precious comments.

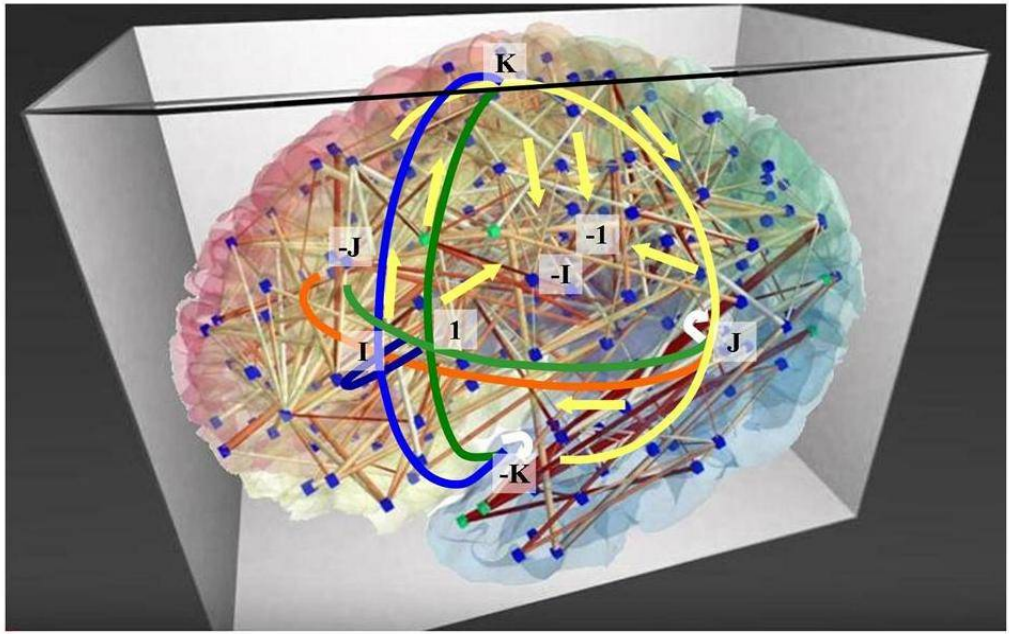

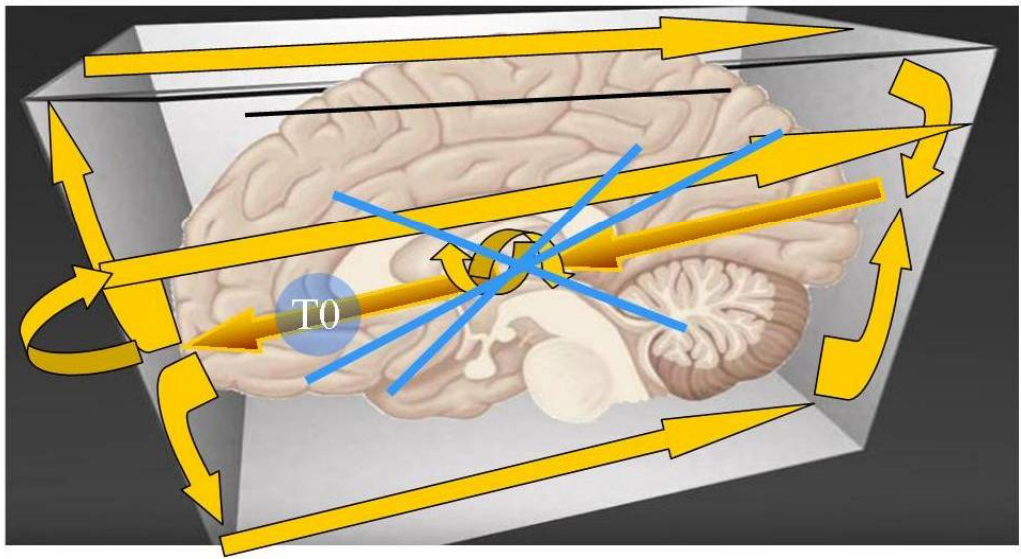

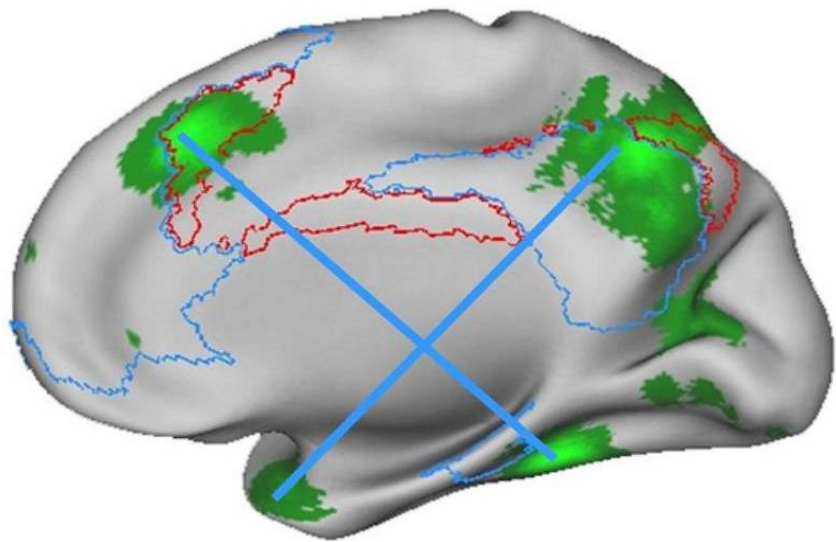

**Figure 2.** The concept of hypersphere in the framework of brain functional activity.
**Figure 2A** shows the brain connectome (both the emispheres are depicted) embedded in the 3D space shown in **Figure 1D**. The position of the hypersphere displayed in Figure is just one of the countless possible: being the glome a functional structure equipped with many rotations and trajectories, it can be placed in different points of the brain surface.
**Figure 2B.** The right brain emisphere is embedded in the 3D space described in **Figures 1D and 2A**. The orange arrows show the 3D projections, in case the brain was located in a 4-ball. The red-orange arrow shows the trajectory of the main stream of the Clifford torus in this case. We displayed just the trajectory from right to left; however, also the opposite trajectory, from left to right, and countless others, can be exploited by the torus during its movements in 4D. The small circle labelled T0 depicts one of the possible starting points, the first activated cortical zone. The nomenclature is borrowed from **Figure 1D**. The blue lines predict the simultaneously activated antipodal points, according to the dictates of the Borsuk Ulam Theorem.
**Figure 2C** depicts a real pattern of fMRI temporal activation. Significant meta-analytic clusters associated with mind-wandering and related spontaneous thought processes (green clusters) juxtaposed with outlines of the default mode network (blue) and the frontoparietal control network (modified from *29*). We can correctly identify the predicted antipodal points (blue lines). Given one point (a brain signal), there is a second point (another brain signal) at the opposite end of a straight line segment connecting them. Other patterns ascribable to the Borsuk Ulam Theorem are available in **Figure 3**.

---------------------------------------------------------------------------------------------------------------------------------

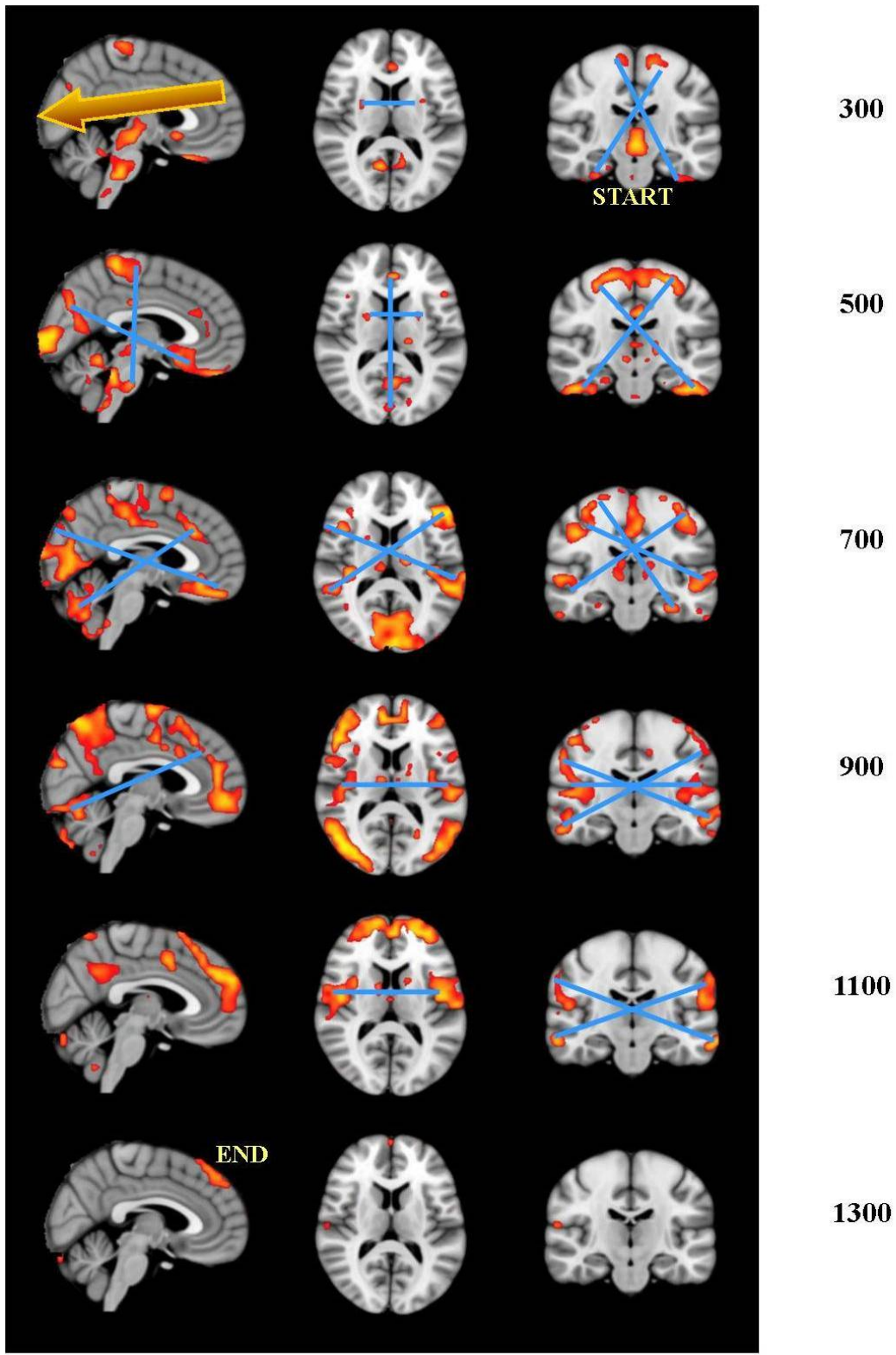

**Figure 3.** Video frames showing "lag threads" computed from real BOLD resting state rs-fMRI data in a group of 688 subjects, obtained from the Harvard-MGH Brain Genomics Superstruct Project (modified from *5*). Note the widely diffused presence of BUT hallmarks (blue lines) at different times. The times are expressed in milliseconds.